
\magnification=\magstep1
\hoffset=0.0 true cm
\vsize=23.5 true cm
\hsize=17.0 true cm

\baselineskip=20pt
\parskip=0pt

\parindent=22pt
\raggedbottom

\font\medrm=cmr10 scaled \magstep1
\font\bigbf=cmb10 scaled \magstep3

\def\pp{\noindent\parshape 2 0.0 truecm 17.0 truecm 1.0 truecm 16.0 truecm}

\def\etal{{\frenchspacing\it et al.} }
\def\lsim{\hbox{ \rlap{\raise 0.425ex\hbox{$<$}}\lower 0.65ex\hbox{$\sim$} }}
\def\gsim{\hbox{ \rlap{\raise 0.425ex\hbox{$>$}}\lower 0.65ex\hbox{$\sim$} }}
\def\kms{{\frenchspacing km s$^{-1}$} }

\def\cgs{{\frenchspacing erg cm$^{-2}$ s$^{-1}$ } }
\def\esa{{\frenchspacing erg s$^{-1}$ \AA$^{-1}$ } }
\def\jnu{{\frenchspacing erg cm$^{-2}$ s$^{-1}$ Hz$^{-1}$ sr$^{-1}$ } }
\def\sbr{{\frenchspacing erg cm$^{-2}$ s$^{-1}$ arcsec$^{-2}$ } }


\null\vskip 4.0 truecm

\centerline{\bigbf Identification of a Galaxy }

\centerline{\bigbf Responsible for a High-Redshift }

\centerline{\bigbf Damped Ly$\alpha$ Absorption System } 
\bigskip
\bigskip

\centerline{\medrm S.~G.~Djorgovski$^\star$, M.~A.~Pahre$^\star$,
J.~Bechtold$^\dagger$, and R.~Elston$^\ddagger$}

\bigskip
\bigskip
\bigskip

\pp $^\star$ Palomar Observatory, MS 105-24, Caltech, Pasadena, CA 91125, USA.

\pp $^\dagger$ Steward Observatory, University of Arizona, Tucson, AZ 85721, USA.

\pp $^\ddagger$ Cerro Tololo Interamerican Observatory, NOAO, Casilla 603,
La Serena, Chile.

\vfill
\centerline{To Appear in {\it Nature}}
\bigskip
\bigskip
\centerline{Received: ...........................................}
\bigskip
\centerline{Accepted: ...........................................}

\vfill
\eject

\baselineskip=12pt
\parskip=5pt

\bf 

Galaxies believed to be responsible for damped Ly$\alpha$ absorption (DLA)
systems in the spectra of high-redshift quasars represent a viable population
of progenitors of normal disk galaxies$^{1}$.  They appear to contain a
substantial fraction of the baryons known to exist in normal galaxies
today$^{2,3}$.  Here we report on the detection of an object, designated DLA
2233+131, responsible for a previously known DLA system at $z_{abs} = 3.150$
[ref. 4] in the spectrum of a quasar 2233+131 ($z_{QSO} = 3.295$)$^{5}$. 
This is the first unambiguous detection of a DLA galaxy, in both emission line
and stellar continuum.  Its properties correspond closely to what may be
expected from a young disk galaxy in the early stages of formation, with no
sign of an active nucleus.  This gives a strong support to the idea that DLA
systems represent a population of young galaxies at high redshifts. 

\bigskip
\rm

Despite considerable efforts, no direct counterpart of a DLA galaxy has been
unambiguously detected so far$^{6,7,8,9}$.  Clustered companions of DLA
objects, all containing AGN, or quasar companions which may be responsible for
some associated absorption have been seen, but no normal, isolated DLA systems
themselves$^{10,11,12,13,14}$.  The closest may be the objects apparently
associated with a quasar PKS 0528--250, with $z_{abs} = 2.811 ~>~ z_{QSO,em} =
2.77$, but given their association with the quasar, their nature and the source
of ionization remain uncertain$^{13,14}$. 

Our data were obtained on the night of UT 1995 September 28, using the Low
Resolution Imaging Spectrograph (LRIS) instrument$^{15}$ at the W.M.~Keck 10-m
telescope on Mauna Kea, Hawaii.  The object was found during observations of a
candidate for a different DLA system (at $z_{abs} = 2.551$)$^{4}$ in the same
field, which will be reported elsewhere.  It was also selected independently as
a DLA candidate on the basis of its broad-band colors by Steidel \etal$^{16}$,
which was unknown to us at the time.  We obtained direct images of the field in
the Cousins $V$ and $R$ bands, and long-slit spectra using low and moderate
resolution gratings, providing spectroscopic resolutions of $FWHM \approx 10$
and 3.5 \AA, respectively.  The seeing was $FWHM \approx 0.7$ arcsec.  Since no
photometric standards were observed that night, $V$ and $R$ images of the same
field were obtained in photometric conditions at the Palomar 60-inch telescope
on the night of UT 1995 October 17, and used to establish the photometric zero
point for the Keck images.  All data were reduced using standard techniques. 

The object is identified with a galaxy 2.3 arcsec away in projection from the
quasar, in $PA = 159^\circ$ (Figure 1).  Its magnitudes are $V = 25.1 \pm 0.2$
and $R = 24.8 \pm 0.2$, in an excellent agreement with the photometry by
Steidel \etal$^{16}$.  The optical spectrum blueward of the Lyman break is thus
nearly flat, with $\langle F_\nu \rangle \approx 0.35 ~\mu$Jy, typical of
actively star forming galaxies.  For the quasar itself we obtain $V = 18.29$
and $R = 18.15$. 

Our spectra show a prominent Ly$\alpha$ line emission at $z_{em} = 3.1530 \pm
0.0003$, corresponding to the restframe velocity difference of $\Delta V = 209$
\kms from the absorption system (Figures 2 and 3).  The central wavelength of
the emission line may be affected by absorption in the ambient gas, which could
modify its observed redshift.  The observed line flux is $F_{1216} = (6.4 \pm
1.2) \times 10^{-17}$ \cgs, where flux zero point uncertainties dominate the
error.  While no other lines are detected, the close proximity to the known
absorber redshift and the consistent colors and magnitudes leave little doubt
as to the redshift identification.  There are no high-ionisation lines detected
(e.g., N V 1240 \AA, or C IV 1549 \AA, which would indicate the presence of an
AGN) with 1-$\sigma$ upper limits of $\sim 10^{-18}$ \cgs. 

Subsequent infrared observations were obtained at the Keck and the Kitt Peak
4-m telescope.  Our measurements are consistent with the estimate$^{17}$ of 
$K \approx 22 \pm 0.3$ mag.  We have an upper limit to the [O III] 5007 \AA\
line flux of $F_{5007} < 10^{-16}$ \cgs, which is consistent from expectations
based on our measurement of the Ly$\alpha$ line flux and simple photoionisation
models.  This suggests that the object is not highly reddened. 

In order to estimate the physical properties of the object, we assume a
standard Friedman model cosmology with $H_0 = 75$ \kms Mpc$^{-1}$ and $\Omega_0
= 0.2$ (our principal conclusions are not sensitive to the exact choice of
these parameters).  This gives the look-back time of 83\% of the age of the
universe at this redshift.  Assuming the onset of galaxy formation at $z \sim
5$, this object is only $\sim 7 \times 10^8$ yr old, which corresponds to only
a few free-fall times for a normal galaxy.  This, and the very blue continuum
implied by our photometry, suggests that DLA 2233+131 must be young.  For our
assumed cosmology, the distance modulus is $(m-M) = 47.11$, the angular size
distance (appropriate for converting observed angular separations into the
restframe linear sizes) is $1.96 \times 10^{28}$ cm, and the luminosity
distance (appropriate for converting observed fluxes into the restframe
luminosities) is $8.14 \times 10^{28}$ cm. 

At this point we cannot be sure if the observed emission is coming from a
young galaxy's disk, or its bulge, or both.  Deep Hubble Space Telescope
images may help resolve this question.  This uncertainty should be borne in
mind in the discussion that follows. 

The projected separation of the detected galaxy and the quasar (i.e., its
absorber portion) is then 17.2 kpc, comparable to the sizes of normal galaxy
disks.  The velocity field, if indeed due to rotation, is comparable to the
rotation curve amplitudes of normal spirals, for reasonable projection angles.
Moreover, the Ly$\alpha$ line shows a red wing (Figure 3), as may be expected
from lines of sight intersecting differentially rotating galaxy
disks$^{18,19}$.  The implied dynamical mass is 
$$ M_{dyn} ~\geq ~{{V^2 R}\over{G}} ~= ~1.86 \times 10^{11} ~M_\odot
~\left( {{V}\over{\rm 200 ~km~s^{-1}}} \right) ^2 
~\left( {{R}\over{\rm 20 ~kpc}} \right) $$
\noindent
(a lower limit due to the unknown projection effects).  From the observed
column density of the neutral hydrogen, $N_{H I} = 1.0 \times 10^{20}$
cm$^{-2}$ [ref. 4], we can estimate the gas mass of 
$$ M_{H I} ~\sim ~\pi ~\langle R \rangle ^2 ~N_{H I} ~m_p
~\approx 1.0 \times 10^9 ~M_\odot
~\left( {{R}\over{\rm 20 ~kpc}} \right) ^2 $$
\noindent
These values are typical of normal galaxy disks today.  They are only meant to
be indicative, given the uncertainties about the morphology of the object and
the projection effects.

The implied Ly$\alpha$ line luminosity is $L_{1216} = 5.3 \times 10^{42}$ erg/s.
Estimates of conversion of the Ly$\alpha$ line luminosity to the implied
unobscured star formation rate ($SFR$) are in the range 
$L_{1216} = (7 \pm 4) \times 10^{41}$ erg/s for 
$SFR = 1 M_\odot$\thinspace yr$^{-1}$, depending
on the stellar initial mass function (IMF)$^{20,21}$.  
We thus estimate the unobscured $SFR$ in DLA 2233+131 to be $SFR
= 7.5 ^{+10} _{-3} ~M_\odot$\thinspace yr$^{-1}$.
An independent estimate of the $SFR$ can be obtained from the restframe 
continuum luminosity at 1500\AA.  For our assumed cosmology, 
$P_{1500} = 9.4 \times 10^{40}$\esa.
Population synthesis models$^{22}$ for a constant 
$SFR = 1 M_\odot$\thinspace yr$^{-1}$, assuming a Salpeter IMF, predict
$P_{1500} \approx 1.5 \times 10^{40}$\esa, but plausible variations in the
IMF slope can change that number by a factor of 3.  With this conversion,
we derive $SFR \approx 6.4 ~M_\odot$\thinspace yr$^{-1}$, in a good agreement
with our estimate from the Ly$\alpha$ line.

This $SFR$, only a few times higher than in most spiral galaxies today, is an
order of magnitude less than the observed star formation rates in
ultraluminous IRAS galaxies, or expected rates in proto-ellipticals, but it is
comparable to what may be expected from a gradually forming, young disk. This
may be a lower limit, if some line emission is absorbed by the gas or dust.  We
derive the restframe equivalent width of 37\AA, less than what most models
would predict$^{20,21}$ for our estimated $SFR$, which can be explained by a
very modest amount of absorption, or the effect of superposed stellar
Ly$\alpha$ absorption. 

Some of the observed Ly$\alpha$ emission may be due to the ionization of the
H I cloud by the metagalactic UV flux, which at this redshift is estimated to
be in the range $J_\nu \approx (1.5 \pm 0.5) \times 10^{-21}$ \jnu [ref. 23].
Following Wolfe \etal $^{6}$, at $z = 3.153$ this implies the observed
Ly$\alpha$ surface brightness of $(1.8 \pm 0.6) \times 10^{-19}$ \sbr, which
is a negligible fraction of the observed line flux.

We can also extrapolate the observed continuum flux to the restframe $B$ band.
Assuming a flat spectrum, we obtain for the observed flux 
$F_\nu (B_{rest}) = 0.36 ~\mu$Jy, 
and for a power law spectrum $F_\nu \sim \nu ^\alpha$, with $\alpha = -0.4$,
the reddest spectrum compatible with our data, 
$F_\nu (B_{rest}) = 0.53 ~\mu$Jy.  In the assumed cosmology, these correspond
to the absolute magnitudes $M_B = -20.33$ and $-20.75$, respectively.  For
comparison, an $L_*$ galaxy today has an $M_B = -20.4$ for the same cosmology.
Thus, the optical luminosity of DLA 2233+131 corresponds to that of normal,
evolved disk galaxies today, and is at least two orders of magnitude lower than
that of typical quasars at that redshift.  Given its higher star formation
rate, we conclude that it yet has to make most of its stars. 

This discovery should then be considered in the context of searches for
protogalaxies.  Despite considerable efforts, no obvious population of
progenitors of normal galaxies has been found in the past$^{24,25,26,27}$. 
One promising technique is to select high-redshift galaxies through their
continuum colors, using the Lyman discontinuity at 912\AA$^{16,17}$.  Most
searches to date have concentrated on emission-line objects, primarily the
Ly$\alpha$ line redshifted in the optical window, or the hydrogen Balmer 
lines or nebular oxygen lines redshifted into the 
near-infrared$^{20,24,25,26,27,28,29}$.  Many interesting high-redshift
galaxies have been found in this way, but essentially all of them either
contain active nuclei (AGN), or are located in a close proximity of one, and
thus may be powered by reprocessed AGN radiation, rather than star formation. 
It is still possible that many or all of the known high-redshift AGN are
situated in young galaxies. 

DLA systems represent an already well-established, {\it large population} of
high-redshift objects for which the confusion with AGN does not arise.  They
have been proposed as likely progenitors of normal disk galaxies$^{1}$, on the
basis of their properties as inferred from the absorption signatures alone.
In support of this idea, the properties of DLA 2233+131, a field galaxy now 
clearly identified with a DLA system, are exactly what may be expected of a
young, still forming disk galaxy at a high redshift.  This object may
eventually evolve into a normal spiral galaxy not very different from our own.
This is also fully consistent with the interpretation of Lyman break objects at
comparable redshifts as a population of proto-bulges$^{17}$.  The populations
of DLA systems, of which DLA 2233+131 may be representative, and of Lyman break
objects may overlap considerably, and they may be tentatively identified as
progenitors of typical normal galaxies today.

\bigskip
\bigskip
\bigskip

\parskip=0pt

\noindent{\bf References:}
\medskip

\pp 1.~ Wolfe, A. M. 
{\it Ann.~New York Acad.~Sci.} {\bf 688}, 281-296 (1993).

\pp 2.~ Lanzetta, K. M., Wolfe, A. M., \& Turnshek, D. A. 
{\it Astrophys. J.} {\bf 440}, 435-457 (1995).

\pp 3.~ Storrie-Lombardi, L. J., McMahon, R. G., Irwin, M. J., \& Hazard, C.
{\it Astrophys. J.} 427, L13-L16 (1994).

\pp 4.~ Lu, L., Wolfe, A. M., Turnshek, D. A., \& Lanzetta, K. M. 
{\it Astrophys. J. Suppl.} {\bf 84}, 1-38 (1993).

\pp 5.~ Crampton, D., Schade, D., \& Cowley, A. P. 
{\it Astr. J.} {\bf 90}, 987-997 (1985).

\pp 6.~ Wolfe, A. M., Turnshek, D. A., Lanzetta, K. M., \& Oke, J. B. 
{\it Astrophys. J.} {\bf 385}, 151-172 (1992).

\pp 7.~ Elston, R., Bechtold, J., Lowenthal, J. D., \& Rieke, M. 
{\it Astrophys. J.} {\bf 373}, L39-L42 (1991).

\pp 8.~ Hu, E. M., Songaila, A., Cowie, L. L., \& Hodapp, K.-W. 
{\it Astrophys. J.} {\bf 419}, L13-L16 (1993).

\pp 9.~ Lowenthal, J. D., Hogan, C. J., Green, R. F., Woodgate, B. E., Caulet,
A., Brown, L., \& Bechtold, J. 
{\it Astrophys. J.} {\bf 451}, 484-497 (1995).

\pp 10. Lowenthal, J. D., Hogan, C. J., Green, R. F., Caulet, A., Woodgate, B.
E., Brown, L., \& Foltz C. B. 
{\it Astrophys. J.} {\bf 377}, L73-L77 (1991).

\pp 11. Macchetto, F., Lipari, S., Giavalisco, M., Turnshek, D. A., \& Sparks,
W. B. 
{\it Astrophys. J.} {\bf 404}, 511-520 (1993).

\pp 12. Francis, P. J., \etal
{\it Astrophys. J.} {\bf 457}, 490-499 (1996).

\pp 13. M{\o}ller, P., \& Warren, S. J. 
{\it Astron. Astrophys.} {\bf 270}, 43-52 (1993).

\pp 14. M{\o}ller, P., \& Warren, S. J. 
{\it Astron. Astrophys.} in press (1996).

\pp 15. Oke, J. B., Cohen, J. G., \etal 
{\it Publs astr. Soc. Pacif.} {\bf 107}, 375-385 (1995).

\pp 16. Steidel, C. C., Pettini, M., \& Hamilton, D. 
{\it Astr. J.} {\bf 110}, 2519-2536 (1995).

\pp 17. Steidel, C. C., Giavalisco, M., Pettini, M., Dickinson, M., \& 
Adelberger, K. 
{\it Astrophys. J.} {\bf 462}, L17-L21 (1996).

\pp 18. Lanzetta, K. M., \& Bowen, D. V. 
{\it Astrophys. J.} {\bf 391}, 48-72 (1992).

\pp 19. Wolfe, A. M., Fan, X.-M., Tytler, D., Vogt, S. S., Keane, M. J.,
\& Lanzetta, K. M. 
{\it Astrophys. J.} {\bf 435}, L101-L104 (1994).

\pp 20. Thompson, D., Djorgovski, S., \& Trauger, J. 
{\it Astr. J.} {\bf 110}, 963-981 (1995).

\pp 21. Charlot, S., \& Fall, S. M. 
{\it Astrophys. J.} {\bf 415}, 580-588 (1993).

\pp 22. Leitherer, C., Robert, C., \& Heckman, T. M. 
{\it Astrophys. J. Suppl.} {\bf 99}, 173-187 (1995).

\pp 23. Haardt, F., \& Madau, P.
{\it Astrophys. J.} {\bf 461}, 20-37 (1996).

\pp 24. Pritchet, C. J. 
{\it Publs astr. Soc. Pacif.} {\bf 106}, 1052-1067 (1994).

\pp 25. Djorgovski, S., \& Thompson, D. 
in {\sl The Stellar Populations of Galaxies}, 
(eds B. Barbuy \& A. Renzini), IAU Symp.~149, 337-347
(Kluwer, Dordrecht, 1992).

\pp 26. Djorgovski, S. 
in {\sl New Light on Galaxy Evolution}, IAU Symp.~171, 
(eds R.~Bender \& R. Davies), p.~277
(Kluwer, Dordrecht, 1996).

\pp 27. Djorgovski, S. 
in {\sl Cosmology and Large-Scale Structure in the Universe}, 
(ed. R. de Carvalho), {\it Astr. Soc. Pacif. Conf. Ser.} {\bf 24}, 73-95 (1992).

\pp 28. Thompson, D., \& Djorgovski, S. 
{\it Astr. J.} {\bf 110}, 982-993 (1995).

\pp 29. Pahre, M. A., \& Djorgovski, S. 
{\it Astrophys. J.} {\bf 449}, L1-L4 (1995).

\bigskip
\bigskip
\noindent{\sl Acknowledgements:}~~ 
This work was based in part on the observations obtained at the W.~M.~Keck
Observatory, which is operated jointly by the California Institute of
Technology and the University of California.  We thank the staffs of Keck and
Palomar observatories for their expert help during our observing runs, and to
Drs. C.~Steidel, L.~Lu, and M.~Rauch for useful conversations.  SGD
acknowledges a support from the US National Science Foundation, and the
Bressler Foundation.

\vfill

\centerline{\bf Figure Captions:}

\bigskip
\noindent{\bf Figure 1:}~~
Finding charts for the field.  The top image is a 3 arcmin square, from $V$ and
$R$ band images obtained at the Palomar 60-inch telescope.  The quasar is
marked with the arrow; its coordinates are: $\alpha = 22^h 33^m 51.1^s$,
$\delta = +13^\circ 10^\prime 46^{\prime\prime}$ (B1950 equinox).  The
bottom image is a 40 arcsec square, from $R$ band images obtained at the Keck
telescope.  The DLA galaxy is indicated with the arrow.  In both images north
is up and east to the left, and the intensity contours are spaced
logarithmically.

\bigskip
\noindent{\bf Figure 2:}~~
Intermediate-resolution spectra of the quasar (QSO, top) and the damped
Ly$\alpha$ galaxy (DLA, bottom), obtained at the Keck telescope.  Absorption
features due to the DLA galaxy are marked in the spectrum of the quasar; in
addition, Si II 1260 line is present in absorption at $\lambda = 5231$\AA, but
it may be confused with the associated Ly$\alpha$ absorption in the quasar
spectrum itself.  Strong Ly$\alpha$ emission is seen in the spectrum of the DLA
galaxy.  Note the complete absence of high-ionisation lines of N V 1240 and C
IV 1549 (marked at the appropriate locations), suggesting that there is no
detectable active nucleus in this object.  The glitch at 5577\AA\ is due to the
poor night sky line subtraction.  The continuum, but not the Ly$\alpha$ line,
is oversubtracted in this spectrum, due to the presence of a much brighter QSO
nearby; the true continuum level should be around 0.35 $\mu$Jy.

\bigskip
\noindent{\bf Figure 3:}~~
The top panel shows a zoom-in on the spectroscopic frame, showing the
Ly$\alpha$ line emission from the DLA galaxy, and the associated absorption in
the quasar spectrum.  Their central redshifts are indicated, and the
corresponding restframe velocity difference is only 209 \kms, comparable to
rotation curve amplitudes of normal disk galaxies.  The projected angular and
velocity scales are indicated.  Note the red wing in the Ly$\alpha$ line,
which is better seen in the bottom panel, showing a zoom-in on the extracted
spectrum.  This is a characteristic line shape expected from differentially
rotating galaxy disks.

\eject

\end